\documentclass[aps,preprint,preprintnumbers,amsmath,amssymb,nofootinbib]{revtex4}
\usepackage{amssymb}
\usepackage{lscape}
\usepackage{amsmath}
\usepackage{rotating}
\usepackage{bm}
\usepackage{graphicx}
\usepackage{epsfig}
\begin{document}
\title {WIMT in Gullstr\"{a}nd-Painlev\'{e} and Reissner-Nordstr\"{o}m metrics: induced Stable Gravito-magnetic Monopoles.}
\author{$^{2}$
Jes\'us Mart\'{\i}n Romero\footnote{E-mail address:
jesusromero@conicet.gov.ar}, $^{1,2}$ Mauricio Bellini
\footnote{E-mail address: mbellini@mdp.edu.ar} }
\address{$^1$ Departamento de F\'isica, Facultad de Ciencias Exactas y
Naturales, Universidad Nacional de Mar del Plata, Funes 3350, C.P.
7600, Mar del Plata, Argentina.\\
$^2$ Instituto de Investigaciones F\'{\i}sicas de Mar del Plata (IFIMAR), \\
Consejo Nacional de Investigaciones Cient\'ificas y T\'ecnicas
(CONICET), Mar del Plata, Argentina.}

\begin{abstract}
The aim of this work is to apply WIMT to Gullstr\"{a}nd-Painlev\'{e} and Reissner-Nordstr\"{o}m metrics in the framework of Weitzeb\"{o}ck Induced Matter Theory (WIMT). This is a newly developed method that extends Induced Matter Theory from a curved 5D manifold using the Weitzeb\"{o}ck's geometry, using the fact that the Riemann-Weitzeb\"{o}ck curvature tensor is always null. We obtain the presence of currents whose interpretation can lead to the presence of stable gravito-magnetic monopoles.
\end{abstract}
\maketitle

\section*{Introduction}
In a previous paper \cite{intro1} we incorporated the  Weitzeb\"{o}ck's geometry (using its characteristic connections) into the treatment of extended Induced Matter Theory(IMT) {\cite{intro2}. This theory is based on the assumption that ordinary matter and physical fields that we can observe in our $4D$ universe can be geometrically obtained from a $5D$ space-time which is at least Ricci-flat (in the sense of Levi-Civita connections). We are interested in the cases where the extra dimension is non-compact \footnote{In contrast with the Kaluza-Klein Theory (KK) in which is assumed that the extra dimension is compact, cyclic and with a small radius. }, and we define a physical vacuum supported by the Ricci-flatness condition \cite{intro3}. This theory is funded in the Campbell-Magaard embedding theorem \cite{intro4} as a particular case in which $^{(5D)}R_{ab}=\lambda\, ^{(5D)}g_{ab}$ \footnote{In our convention indices $"a,b,c,...,h"$ and $"A,B,C,...,H"$ are related to $5D$ space-time and run from $1$ to $5$. Greek indices $"\alpha,\beta,\gamma,..."$ are related to $4D$ space-time and run from $1$ to $4$, and indices $"i,j,k,..."$ are only spatial and run from $2$ to $4$. We choose our first index related to time coordinate and the last index in the $5D$ space-time is due to extra coordinate associated to extra dimension. }, with $\lambda=0$. We must remark that the $\lambda=0$ condition implies Ricci-flatness and this makes possible to define a 5D apparent vacuum that determines the equations of motion for the fields of the theory. Such equations are given by $^{(5D)}R_{ab}=0$. The aim of WIMT is to use an alternative description, based in the Weitzenb\"{o}ck geometry, to apply a IMT-like formalism for any $5D$ space-time even if this is not flat in the Riemannian sense with Levy-Civita (LC) connections. The central idea of WIMT is to convert the geometrical description of the problem from a non-flat (but torsion-less) Riemannian  with LC connections, into a Weitzenb\"{o}ck's geometrical description with a Riemann-Weitzenb\"{o}ck null tensor \begin{footnote}{What we refer to as Riemann-Weitzenb\"{o}ck tensor is the Riemann curvature tensor expressed in terms of Weitzenb\"{o}ck connections. In a coordinate (holonomic) basis the Riemann-Weitzenb\"{o}ck curvature takes the form
\begin{eqnarray}\nonumber ^{(W)}R^a_{bcd}&=&^{(W)}\Gamma^a_{dc\,,\,b} -^{(W)}\Gamma^a_{db\,,\,c} +
\,^{(W)}\Gamma^n_{dc} \,^{(W)}\Gamma^a_{nb}-\,^{(W)}\Gamma^n_{db}\,^{(W)}\Gamma^a_{nc}. \end{eqnarray}
The label $^{(W)}$ identifies the geometric objects expressed in the Weitzenb\"{o}ck representation and $^{(W)}\Gamma^a_{dc\,,\,b}$ terms are the ordinary partial derivatives of the Weitzenb\"{o}ck connection $^{(W)}\Gamma^a_{dc}$. We assume the Einstein summation convention for repeated indices.}\end{footnote}, but with non-zero torsion. In this case we can use the IMT tools to induce $4D$ effective space-time dynamics making a foliation over the Riemann-Weitzenb\"{o}ck flat $5D$ space-time (this is the vacuum in the Weitzenb\"{o}ck sense) \begin{footnote}{We set the Weitzenb\"{o}ck vacuum as a $5D$ space-time with null Riemann curvature tensor in Weitzenb\"{o}ck geometry, \begin{equation}\nonumber ^{(W)}R^a_{bcd}=0.\end{equation}}\end{footnote}. Once this choice has been made, we can recover the Riemannian description by doing a transformation over the induced tensors (now obtained in terms of Weitzenb\"{o}ck connections) according to $^{(LC)}\Gamma^{a}_{bc}=^{(W)}\Gamma^{a}_{bc}+K^{a}_{bc}$. Here $^{(LC)}\Gamma^{a}_{bc}$ is the symbol denoting the LC connection (the usual Christoffel symbols of the second kind), $^{(W)}\Gamma^{a}_{bc}$ denotes the Weitzenb\"{o}ck connections and $K^{a}_{bc}$ is the contortion, which depends on the
non-metricity and torsion. The shape of the contortion is presented in the next section.

Part of the motivation for the present work comes from two principal ideas:
{\bf i)} WIMT provides a successfully description for the formation of a spherically symmetric body of finite size, and {\bf ii)} the Weitzenb\"{o}ck torsion present in WIMT is a possible geometrical source for the existence of magnetic monopoles.

With respect to the first assertion we can say that already the WIMT was used to address the problem of the formation of a massive compact object\cite{intro1} as a remanent of a collapsing space-time. This object can be thought of as a black hole. The second one has been worked using WIMT in the framework of the gravito-magnetic formalism in a variety of cosmological scenarios\cite{intro6}, with the result that  currents of gravito-magnetic monopoles are strongly linked to the Weitzenb\"{o}ck torsion. With the support of these results, it is generally expected that the density of gravito-magnetic monopoles and charges decay in an accelerated universe due to the strong increasing of the scale factor. Hence, the study of a stationary black hole appears to be an excellent issue to complete our approach to a spherically symmetric object and to try to appreciate the existence of stable gravito-magnetic charges using WIMT. We choose to work in the 5D extension of Gullstr\"{a}nd-Painlev\'{e} (GP) metric, because with this metric is very simple to get the associated basis of pentads and to operate with them. Additionally, but no less important, the GP metric is widely used in the study of BH and their evolution\cite{intro7}, so that it is an interesting topic that deserves further study.

\section{WIMT.}

We shall introduce some basic concepts of the Weitzenb\"{o}ck geometry and we expose some results that we developed in previous works, because these will be tools for obtaining some important results of this article.

\subsection{Weitzenb\"{o}ck geometry and vielbeins}\label{ssw}

In order to revisit the WIMT we shall use some elements of Weitzenb\"{o}ck geometry\cite{wg}. The Weitzenb\"{o}ck connection is
\begin{equation}\label{w00}
  ^{(W)}\Gamma^c_{ab}=\bar{e}^c_N\overrightarrow{e}_b(e_a^N);
\end{equation}
we assume the transformation
\begin{equation}\label{w01}
\overrightarrow{e}_a = e^A_{\,\,a} \overrightarrow{E}_A,
\end{equation}
which represents a transformation from an element $\overrightarrow{E}_A$ (belonging to $\{\overrightarrow{E}_A\}$, an orthonormal Lorentzian basis of the 5D tangent space), into $\overrightarrow{e}_a$ (belonging to $\{\overrightarrow{e}_a\}$, also a basis of 5D tangent space but in general not orthonormal). The elements $e^A_{\,\,a}$ that perform the transformation are known as vielbeins\cite{viel} and fulfill the relations
$e^A_{\,\,a}
\bar{e}^b_{\,\,A} = \delta^b_a$ and $e^A_{\,\,a}
\bar{e}^a_{\,\,B} = \delta^A_B$. The vielbeins provide us with a prescription to transform bases and therefore any tensorial element. For example,
\begin{equation}\label{w02}
g_{ab} = e^A_{\,\,a}e^B_{\,\,b} \eta_{AB},
\end{equation}
where $\eta_{AB}$ is a 5D Minkowski-like metric with signature $(+,-,-,-,-)$, $\{\overrightarrow{E}_A\}$ is an orthonormal Lorentzian basis and $g_{ab}$ is the metric tensor expressed in terms of the basis $\{\overrightarrow{e}_a\}$. This clarifies that $e^A_{\,\,a}e^B_{\,\,b}$ is an object with five "legs" (corresponding to the five components of the space-time), in the $\{\overrightarrow{E}_A\}$ representation for tensorial elements (upper case indices) and the other five "legs" in the $\{\overrightarrow{e}_a\}$ representation (lower case indices). For a general tensor ${T^{A_1...A_n}}_{B_1...B_m}$, which is $n$ times tangent (or contravariant) and $m$ times cotangent (or covariant), we obtain
\begin{equation}\label{w03}
{T^{a_1...a_n}}_{b_1...b_m}=\bar{e}^{a_1}_{A_1}...\bar{e}^{a_n}_{A_n}e^{B_1}_{b_1}...e^{B_m}_{b_m}{T^{A_1...A_n}}_{B_1...B_m}.
\end{equation}
An important property of the Weitzenb\"{o}ck connection $^{(W)}\Gamma^c_{ab}=\bar{e}^c_N\overrightarrow{e}_b(e_a^N)$ is that
\begin{eqnarray}
^{(W)}\nabla_{\overrightarrow{e}_b}\left(\overrightarrow{E}_A\right)&=&\,^{(W)}\nabla_{\overrightarrow{e}_b}\left(e^a_A\overrightarrow{e}_a\right)=\underbrace{e^a_A\left\{^{(W)}\Gamma^c_{ab}
-e^c_N\overrightarrow{e}_b\left(e^N_a\right)\right\}}\overrightarrow{e}_c=0.\nonumber \\
&\,&\,\,\,\,\,\,\,\,\,\,\,\,\,\,\,\,\,\,\,\,\,\,\,\,\,\,\,\,\,\,\,\,\,\,\,\,\,\,\,\,\,\,[^{(W)}\nabla_{\overrightarrow{e}_b}(\overrightarrow{E}_A)]^c=
{[\overrightarrow{E}_A]^c}_{;\,b}=\bar{e}^c_{A;\,b}\label{w04}
\end{eqnarray}
This means that the covariant Weitzenb\"{o}ck derivative of the vielbein is null. The expression (\ref{w00}) characterizes the Weitzenb\"{o}ck connection with lower case indices as a transformation of a trivial null connection with upper case indices. About the trivial upper case index connection we must be more formal and construct the following expression
\begin{eqnarray} ^{(W)}\Gamma^{C}_{BA}\overrightarrow{E}_C &=& \nabla_{\overrightarrow{E}_A}(\overrightarrow{E}_B)=\bar{e}^{a}_{A}\nabla_{\overrightarrow{e}_a}(\bar{e}^{b}_{B}\overrightarrow{e}_b)=\bar{e}^{a}_{A}\overrightarrow{e}_a(\bar{e}^{b}_{B})
\overrightarrow{e}_b+\bar{e}^{a}_{A}\bar{e}^{b}_{B}\underbrace{\nabla_{\overrightarrow{e}_a}(\overrightarrow{e}_b)}\\ \nonumber &\,&\,\,\,\,\,\,\,\,\,\,\,\,\,\,\,\,\,\,\,\,\,\,\,\,\,\,\,\,\,\,\,\,\,\,\,\,\,\,\,\,\,\,\,\,\,\,\,\,\,\,\,\,\,\,\,\,\,\,\,\,\,\,\,\,\,\,\,\,\,\,\,\,\, \,\,\,\,\,\,\,\,\,\,\,\,\,\,\,\,\,\,\,\,\,\,\,\,\,\,\,\,\,\,\,\,\,\,\,\,\,\,\,\,\,\,\,\,\,\,\,\,\,\,\,\,\,\, ^{(W)}\Gamma^{c}_{ba}\overrightarrow{e}_c=\bar{e}^{c}_{N}\overrightarrow{e}_a(e^{N}_b)\overrightarrow{e}_c \\\nonumber &=&\bar{e}^{a}_{A}\{\overrightarrow{e}_a(\bar{e}^{b}_{B})-\bar{e}^{b}_{B}e^{N}_b\overrightarrow{e}_a(\bar{e}^{b}_{N})\}\overrightarrow{e}_b=0,  \label{pie0}
\end{eqnarray}
where we have used $e^{N}_{b}\bar{e}^{b}_B=\delta^{N}_{B}$, $e^{N}_{b}\bar{e}^{c}_N=\delta^{c}_{b}$ and consequently $\overrightarrow{e}_a(e^{N}_{b})\bar{e}^{c}_{N}=-e^{N}_{b}\overrightarrow{e}_a(\bar{e}^{c}_{N})$. The elements $\overrightarrow{E}_C$ are linearly independent so that $^{(W)}\Gamma^{C}_{AB}=0$. In this way the Weitzenb\"{o}ck connection represents the extension of the derivative operator with trivial zero connections in the orthonormal basis to the new basis. In the most general case the Riemann curvature tensor in the Weitzenb\"{o}ck geometry takes the following shape in terms of upper case index connection:
\begin{eqnarray}\label{pie1}
^{(W)}R^A_{BCD}&=&\overrightarrow{E}_B\left(^{(W)}\Gamma^A_{DC}\right) -\overrightarrow{E}_C\left(^{(W)}\Gamma^A_{DB}\right) +
\,^{(W)}\Gamma^N_{DC} \,^{(W)}\Gamma^A_{NB} \\\nonumber
&\,&-^{(W)}\Gamma^N_{DB}\,^{(W)}\Gamma^A_{NC}- C^N_{CB}\,
^{(W)}\Gamma^A_{DN}=0,
\end{eqnarray}
due to each upper case index connection being null: $^{(W)}\Gamma^N_{MO}=0$, as we can see in (\ref{pie0}). The structure coefficients in the expression (\ref{pie1}) are defined by $\left[\overrightarrow{E}_B,\overrightarrow{E}_A\right]=C^C_{AB}\overrightarrow{E}_C$. Therefore, the next expression is trivial:
{\begin{equation}\label{w05}
^{(W)}R^a_{bcd}=\bar{e}^{a}_{A}e^{B}_{b}e^{C}_{c}e^{D}_{d}\,^{(W)}R^A_{BCD}=0.
\end{equation}
This relationship is the key of WIMT. It provides us with a tool to define the 5D Weitzenb\"{o}ck vacuum and make use of the usual philosophy of IMT\footnote{The central ideas of IMT are very didactically developed in\cite{intro01}.}, even in absence of the 5D Riemannian(LC) vacuum\begin{footnote}{We want to clarify what we mean by "Riemannian(LC) vacuum": it means that the Einstein tensor written in terms of LC connections is zero. This is trivial when the Riemann curvature tensor is null.}\end{footnote}.

Finally, the relationships between the LC and Weitzenb\"{o}ck connections are
\begin{equation}\label{w06}
^{(W)}\Gamma^a_{bc} = ^{(lc)}{\Gamma}^a_{bc} - \,^{(W)}K^a_{bc}.
\end{equation}
where the Weitzenb\"{o}ck contortion $^{(W)}K^a_{bc}$, is
\begin{equation}\label{w07}
^{(W)}K^a_{bc}=\frac{g^{ma}}{2}\{g_{bm;c}
+g_{mc;b}-g_{bc;m}\}+\frac{g^{ma}}{2}\{^{(W)}T^n_{cm}g_{bn}+\,^{(W)}T^n_{bm}g_{nc}-\,^{(W)}T^n_{cb}g_{nm}\}.
\end{equation}
In the last expression $g_{ab\,;\,c}=\,^{(W)}Q_{abc}$ is the Weitzenb\"{o}ck non-metricity.
In the absence of such non-metricity we recover the usual shape for contortion, which is totally given in terms of the torsion tensor\cite{intro02}.
 In the present cases we make a start with an orthonormal basis $\{\overrightarrow{E}_A\}$. The Weitzenb\"{o}ck non-metricity is
 \begin{equation}\nonumber ^{(W)}Q_{ABC}=\eta_{AB\,;\,C}=\overrightarrow{E}_{C}(\eta_{AB})-\eta_{NB}\,^{(W)}\Gamma^{N}_{AC}-\eta_{AN}\,^{(W)}\Gamma^{N}_{BC},
 \end{equation}
 which reduces to $^{(W)}Q_{ABC}=\overrightarrow{E}_C(\eta_{AB})$, because of $^{(W)}\Gamma^{A}_{BC}=0$. We must remember that each $\eta_{AB}$ is constant and then
 \begin{equation}\nonumber
 ^{(W)}Q_{ABC}=\overrightarrow{E}_C(\eta_{AB})=0.
 \end{equation}
The last expression is general but in the next section we shall study some case in which this is specially easy to see. We shall work with a holonomic coordinate basis $\{\overrightarrow{e}_a\}$, so that
\begin{equation}\nonumber
^{(W)}Q_{ABC}=\overrightarrow{E}_C(\eta_{AB})=\bar{e}^{c}_C\overrightarrow{e}_{c}(\eta_{AB})=\bar{e}^{c}_C\,\eta_{AB\,,\,c}=0,
\end{equation}
because of $\eta_{AB\,,\,c}=0$. In the last expression $\eta_{AB\,,\,c}=\frac{\partial \eta_{AB}}{\partial x^{c}}$, where the coordinate $x^{c}$ is related to
the $c$-th element of the coordinate basis $\overrightarrow{e}_c$. The Weitzenb\"{o}ck non-metricity in lower case indices is given by $^{(W)}Q_{abc}=e^{A}_{a}e^{B}_{b}e^{C}_{c}\,\,^{(W)}Q_{ABC}=0$. Because of all this, in the present case we must set \begin{equation}\nonumber
^{(W)}K^a_{bc}=\frac{g^{ma}}{2}\{^{(W)}T^n_{cm}g_{bn}+\,^{(W)}T^n_{bm}g_{nc}-\,^{(W)}T^n_{cb}g_{nm}\},
\end{equation}
which is the usual well-known contortion tensor\cite{intro02}. This tensor transforms according to
$^{(W)}K^A_{BC}=e^{A}_{a}\bar{e}^{b}_{B}\bar{e}^{c}_{C}\,^{(W)}K^a_{bc}$.

\subsection{Some results in WIMT}\label{ssr}

We consider a 5D space-time with coordinates $\phi(p)=(t,x,y,z,\psi)_p$ dotted by an inner product described by $Diag(\eta_{AB})=(+1,-1,-1,-1,-1)$, expressed in a non-holonomic basis.
The action on the 5D apparent vacuum is given by the gravito-electromagnetic fields
\begin{eqnarray}\label{apaccionfaraday}\mathcal{S}=\int d^5x\sqrt{|\eta|}\left[\frac{R}{16\,\pi\,G}-\frac{1}{4}
F_{AB}
F^{AB}-\frac{\lambda}{2}\left(A^B_{;\,B}\right)^2\right],\end{eqnarray}
where the penta-vector potential is $\overrightarrow{A}=(\phi,\mathbf{A},\Psi)$, $\phi$ is the electric potential, $\mathbf{A}=(A^x,A^y,A^z)$ is the vector potential, and $\Psi$ is the
gravitational scalar potential. If we assume the Lorentz gauge $A^B_{;\,B}=0$, the equations of motion will be reduced to
\begin{eqnarray}\label{apdina0}
\Box A^K&=&\eta^{BC} A^K_{;\,BC}=0.
\end{eqnarray}
In general, we find that the 5D Faraday tensors expressed in terms of the LC and Weitzenb\"{o}ck versions are given, respectively, by
\begin{eqnarray} ^{(LC)}F_{AB}&=&
\overrightarrow{E}_A(A_B)-\overrightarrow{E}_B(A_A)+A_N C^N_{AB},\\ \nonumber ^{(W)}F_{AB}&=&
\overrightarrow{E}_A(A_B)-\overrightarrow{E}_B(A_A)+A_N(^{(W)}T^N_{BA}+C^N_{AB}),\\ \nonumber ^{(LC)}F^{NB}&=&\,^{(W)}F^{NB}+\eta^{RN}A^PK^B_{PR}-\eta^{RB}A^PK^N_{PR}.
\end{eqnarray}
The 5D-current of gravito-magnetic monopoles is obtained (still in 5D space-time) from the relation $^{(*)}J_{AB}={^{(*)}\mathfrak{F}_{ABC}}^{;C}$.
This is motivated by the symmetry of Maxwell equations\footnote{In a 4D space-time, this symmetry is manifest in the expressions \begin{eqnarray}\label{apmax3}
{F_{\mu\nu}}^{;\nu}&=&^{(e)}J_{\mu},\\ \nonumber
{\mathfrak{F}_{\mu\nu}}^{;\nu}&=&^{(m)}J_{\mu}.
\end{eqnarray}
Here, the labels $e$ and $m$, identify, respectively, the electrical and magnetic quantities according to $^{(e)}\overrightarrow{J}=(\rho_e,\,^{(e)}J_1,\,^{(e)}J_2,\,^{(e)}J_3)$ and $^{(m)}\overrightarrow{J}=(\rho_m,\,^{(m)}J_1,\,^{(m)}J_2,\,^{(m)}J_3)$. Furthermore, $F_{\mu\nu}$ is the Faraday tensor and $\mathfrak{F}_{\mu\nu}$ is its dual\cite{dirac}.}.
Thus, $*$ must be LC or W according to the corresponding derivative operator. Furthermore, $\mathfrak{F}_{ABC}$ is the dual tensor of $F_{AB}$ \begin{footnote}{The dual tensor of a 2-form (characterized in this case by $^{(*)}F_{AB}$) in 5D manifold is a 3-form defined by \begin{equation}\nonumber ^{(*)}\mathfrak{F}_{ABC}=\frac{1}{2}\varepsilon_{ABCDE}\,^{(*)}F^{DE}.\end{equation} Here, $*$ must be LC or W and $\varepsilon_{ABCDE}$ is the antisymmetric Levi-Civita symbol.
We have made use of the fact that $\sqrt{\mid\eta\mid}=1$, being $\eta=Det(\left[\eta\right]_{AB})$ the determinant of metric tensor in the upper case index representation.}\end{footnote}.
The resulting currents, 2-form in the 5D space-time, are
{\footnotesize{\begin{eqnarray}\label{ap6.02}\\ \nonumber
^{(LC)}\underrightarrow{\underrightarrow{J}}
=\frac{\sqrt{|\eta|}}{2}
\varepsilon_{ABCDE}\,\eta^{CF}\,\eta^{DG}\,\eta^{EH}\left[^{(LC)}\underline{d}\left(^{(LC)}\underrightarrow{
\underrightarrow{F}}\right)\right]_{[FGH]}\underrightarrow{E}^A\otimes\underrightarrow{E}^B,\end{eqnarray}}}
\begin{eqnarray}\label{apcosa}\\\nonumber\left[^{(LC)}\underline{d}\left(^{(LC)}
\underrightarrow{\underrightarrow{F}}\right)\right]_{[ABC]}&=&^{(W)}T^N_{AC}\overrightarrow{E}_N(A_B)+^{(W)}T^N_{BA}
\overrightarrow{E}_N(A_C)+^{(W)}T^N_{CB}\overrightarrow{E}_N(A_A)\\\nonumber
&\,&+^{(W)}T^n_{AC}\overrightarrow{E}_B(A_n)
+^{(W)}T^n_{CB}\overrightarrow{E}_A(A_n)+^{(W)}T^n_{BA}\overrightarrow{E}_C(A_n)\\\nonumber
&\,&+\overrightarrow{E}_A(^{(W)}T^n_{CD})A_n+
\overrightarrow{E}_B(^{(W)}T^n_{AC})A_n+\overrightarrow{E}_C(^{(W)}T^n_{AB})A_n.
\end{eqnarray}
The Riemannian (LC) current of gravito-magnetic monopole is originates in the Weitzenb\"{o}ck torsion $^{(W)}T^A_{BC}$\begin{footnote}{We focus our attention on the
Riemannian (LC) current, but is formally easy to relate the Riemannian (LC) and Weitzenb\"{o}ck  currents of the gravito-magnetic monopole, with the expressions
\begin{eqnarray}\label{ap6.03}^{(LC)}J_{AB}-\,^{(W)}J_{AB}=\frac{\sqrt{|\eta|}}{2}\varepsilon_{ABCDE}\frac{1}{4}M^{[CDE]},
\end{eqnarray}
where $M^{[CDE]}=\eta^{CF}\eta^{DG}\eta^{EH}M_{[FGH]}$, and
\begin{eqnarray}\label{ap6.04}\\ \nonumber
M_{[FGH]}&=&\left(A_M\,^{(W)}T^M_{[FG}\right)\,_{;\,H]}-2\,^{(W)}T^N_{[FH|}
\,^{(W)}T^M_{N|G]}A_M-2\,^{(W)}T^N_{[GH}\,^{(W)}T^M_{F]N}A_M\\ \nonumber
&\,&-\,^{(W)}T^N_{[FH|}\overrightarrow{E}_N(A_{|G]})+\,^{(W)}
T^N_{[GH|}\overrightarrow{E}_N(A_{|F]})+\,^{(W)}T^N_{[FH}\overrightarrow{E}_{G]}(A_{N})-\,^{(W)}T^N_{[GH}\overrightarrow{E}_{F]}(A_{N}).\end{eqnarray}}\end{footnote}.
It must be noticed that once we choose the Lorenz gauge condition
$^{(LC)}A^N_{;\,N}=0$, the same condition is preserved in the
Weitzenb\"{o}ck representation: $^{(W)}A^n_{;\,n}=0$.

Once we have established the 5D-currents it is easy to obtain the 4D-induced currents by applying a constant foliation on the fifth extra dimension: $\psi=\psi_0$. The effective resulting 4D
gravito-magnetic current is the 1-form given by
\begin{equation}\label{ssr01}
^{(*)4D}\underrightarrow{J}(\,\,)=\left.\,^{(*)5D}\underrightarrow{\underrightarrow{J}}\left(\overrightarrow{n},\,\, \right)\right|_{\psi=\psi_0},
\end{equation}
so that $*$ must be LC or W and $\overrightarrow{n}$ is a vector pointing to the extra direction. The vector $\overrightarrow{n}$ is normalized and orthogonal to the
4D tangent sub-space associated with the effective 4D sub-manifold. Such orthogonality is realized with respect to the complete 5D metric. All expressions must be evaluated in $\psi=\psi_0$, so that $dx^5\mid_{\psi=\psi_0}=d\psi\mid_{\psi=\psi_0}=0$ and hence $^{(*)4D}\underrightarrow{J}$ must be a cotangent four-vector (i.e., it is a 1-form in the cotangent 4D space-time).

\section{5D Gullstr\"{a}nd-Painlev\'{e} (GP) metric.}

The GP coordinates are commonly presented as a set of coordinates obtained by transformation of the Schwarzschild, or Reissner-Nordstr\"{o}m (RN) ones, in which the static time is replaced by a GP time $t\rightarrow\tilde{t}(t,r)$, preserving the spherical symmetry of the problem\cite{gpmetric00}. We shall use a slightly different method by using the GP cartesian metric and using the vielbein related to an orthonormal basis. In this case we obtain the GP cartesian metric from a normalized one, using a certain vielbein, prescribed in this section. The GP cartesian metric must be obtained from the Schwarzschild or RN BH by doing a space-time transformation. We choose to start with an orthonormal basis $\{\overrightarrow{E}_A\}$, in which
\begin{equation}\label{egp00}
\overrightarrow{E}_A=\bar{e}^{a}_{A}\overrightarrow{e}_a,
\end{equation}
where $\left\{\overrightarrow{e}_a\right\}$ is some coordinate basis. The vielbein is constructed according to
\begin{eqnarray}\label{egp01}
  [\bar{e}^{m}_{M}] &=& \left(
          \begin{array}{ccccc}
            1 & 0 & 0 & 0 & 0 \\
            -{\beta}^{x} & 1 & 0 & 0 & 0 \\
            -{\beta}^{y} & 0 & 1 & 0 & 0 \\
            -{\beta}^{z} & 0 & 0 & 1 & 0 \\
            0 & 0 & 0 & 0 & 1 \\
          \end{array}
        \right),
\end{eqnarray}
and its inverse is
\begin{eqnarray}\label{egp02}
  [e^{M}_{m}] &=& \left(
          \begin{array}{ccccc}
            1 & {\beta}^{x} & {\beta}^{y} & {\beta}^{z} & 0 \\
            0 & 1 & 0 & 0 & 0 \\
            0 & 0 & 1 & 0 & 0 \\
            0 & 0 & 0 & 1 & 0 \\
            0 & 0 & 0 & 0 & 1 \\
          \end{array}
        \right),
\end{eqnarray}
in which $\beta^{i}=\beta x^{i}$, such that $x^{i}$ is the i-th coordinate associated to the coordinate basis $\{\overrightarrow{e}_a\}$, $\beta=\sqrt{\frac{2(M-\frac{Q^2}{2r})}{r^3}}$ is a function, $M$ is a mass parameter, and $Q$ is an electric charge parameter. If $Q=0$ we deal with a Schwarzschild BH, but in the case $Q\neq 0$, we are in presence of a Reissner-N\"{o}rdstrom BH. We must remark that $M-\frac{Q^2}{2r}$ is the gravito-electric charge of the BH, and
\begin{equation}\nonumber
\beta^{i}\beta_{i}=\beta^{I}\beta_{I}=-\left(\beta^2 x^2+\beta^2 y^2+\beta^2 z^2\right)=-\beta^2r^2,
\end{equation}
so that, by definition of $\beta$, we have \begin{equation}\nonumber
\beta^{2}=2\left(M-\frac{Q^2}{2r}\right)\frac{1}{r^3}=-\frac{1}{r^2}\left(\frac{2M}{r}-\frac{Q^2}{r^2}\right).
\end{equation}
Hence
\begin{equation}\nonumber\beta^{i}\beta_{i}=-\left(\frac{2M}{r}-\frac{Q^2}{r^2}\right).
\end{equation}}
The charge of the BH is dominant at small distances, but the mass dominates at long distances. The metric is
\begin{eqnarray}\label{egp03}
dS^2&=& \underbrace{\eta_{MN}\,e^{M}_{m}e^{N}_{n}} dx^m\,dx^n=\eta_{MN} \underbrace{e^{M}_{m}dx^m}\underbrace{e^{N}_{n}dx^n},\nonumber \\  &\,&\,\,\,\,\,\,\,\, g_{mn} \,\,\,\,\,\,\,\,\,\,\,\,\,\,\,\,\,\,\,\,\,\,\,\,\,\,\,\,\,\,\,\,\,\,\,\,\,\,\,\,\,\,\,\,\,\,\,\,\,\,\,\,\,\,\,\,\,\,\,\, E^M\,\,\,\,\,\,\,\,\,\,\,\,E^N  \nonumber \\ &=&dt^2-\delta_{ij}(dx^i-\beta^i dt)(dx^j-\beta^j dt)-d\psi^2.                                                                                                                                           \end{eqnarray}
In the last expression $E^N=E^{N}_{n}dx^n$ is an element belonging to a basis for the cotangent space, which fulfills the duality relation $E^{N}(\overrightarrow{E}_M)=\delta^{N}_{M}$.
Furthermore, $g_{mn}$ is the corresponding GP metric in cartesian 5D GP coordinates, such that $x^1=t$, $x^2=x$, $x^3=y$, $x^4=z$ and $x^5=\psi$\footnote{For a most extensive treatment the reader
could see the notes of Andrew Hamilton (developed in 4D)\cite{gpCASA}.}.

Now we must calculate the structure coefficients for the basis (\ref{egp00}). According to the
definition $\left[\overrightarrow{E}_B,\overrightarrow{E}_A\right]=C^C_{AB}\overrightarrow{E}_C$\cite{RG}, we obtain
\begin{equation}\label{egp04}
  C^{C}_{AB}=\bar{e}^{m}_{B}\frac{\partial\bar{e}^{l}_{A}}{\partial x^m}e^{C}_{l}-\bar{e}^{m}_{A}\frac{\partial \bar{e}^{l}_{B}}{\partial x^m}e^{C}_{l}.
\end{equation}
The non-zero elements are $C^1_{21}=-C^1_{12}$, $C^3_{32}=-C^3_{23}=-\frac{\sqrt{1+\beta^i\beta_i}}{\sqrt{x^2+y^2+z^2}}=-\frac{\sqrt{f(r)}}{r}$,  $C^4_{42}=-C^4_{24}=-\frac{\sqrt{1+\beta^i\beta_i}}{\sqrt{x^2+y^2+z^2}}=-\frac{\sqrt{f(r)}}{r}$ and $C^4_{43}=-C^4_{24}=-\frac{z}{\sqrt{x^2+y^2}\sqrt{x^2+y^2+z^2}}=-\frac{\cos(\theta)}{r \sin(\theta)}$ are expressed in GP, or in the usual BH coordinates.
Here, $f(r)=1-\frac{2M}{r}+\frac{Q^2}{r^2}$ for a Reissner-Nordstr\"{o}m BH. In the case with $Q=0$ one obtains a Schwarzschild BH. On the other hand,
if we take the expression (\ref{egp04}), will be easily shown that $C^{c}_{ab}=0$, which means that $\{\overrightarrow{e}_{a}\}$ is a coordinate or holonomic basis for 5D tangent space. It is very interesting to notice that in the upper case index representation, the torsion tensor is numerically equal to the structure coefficients, due to the fact that $^{(W)}T^{A}_{BC}=\,^{(W)}\Gamma^{A}_{CB}-^{(W)}\Gamma^{A}_{BC}+C^{A}_{BC}$. Since the Weitzenb\"{o}ck connection extends the trivial null connection in the upper case index basis, we obtain that $^{(W)}\Gamma^{A}_{CB}=0$, and hence
\begin{equation}\label{egp05}
^{(W)}T^{A}_{BC}=C^{A}_{BC}=-C^{A}_{CB},
\end{equation}
which is surprising because $^{(W)}T^{A}_{BC}$ is an intrinsic geometrical property of the manifold and $C^{A}_{BC}$ is a coefficient depending on a circumstantially selected basis. In fact the expression (\ref{egp05}) is not strange and is due to the selection of the Weitzenb\"{o}ck connection and particularly to the geometrical properties to extend a derivative operator linked to a trivial null connection in the upper case index, to a non trivial connection in the lower case index basis.

\section{Gravito-magnetic monopoles from WIMT in GP and RN metrics.}

If we take the structure coefficients (\ref{egp04}) and we employ the non-zero elements in (\ref{egp05}), we obtain the Weitzenb\"{o}ck torsion and we
could use it in the expressions (\ref{ap6.02}) and (\ref{apcosa}), in order to write the 5D LC 2-form currents in upper case index basis.
After it, we can express such 2-form in the lower case index representation with the transformation (\ref{w03}) adapted to a 2 times cotangent object [as we doing in (\ref{w02}) for the
metric tensor], with the vielbein (\ref{egp02}). Notice that we have a 5D LC 2-form current expressed in terms of a Weitzenb\"{o}ck torsion with lower case indices and vielbeins. Hence, we can make a constant foliation getting the effective 4D current from (\ref{ssr01}). Finally we obtain the expression for the effective 4D current, which in the usual BH coordinates takes the form
\begin{eqnarray}\label{current0}
&\,&\\\nonumber ^{(LC)(m)}J_1\mid_{\psi=\psi_0}&=&\rho_m=\\\nonumber & = & K\left\{-\frac{\sqrt{f(r)}}{r}\frac{\partial A_3}{\partial\theta}+\frac{\sqrt{f(r)}}{r \sin(\theta)}\frac{\partial A_2}{\partial\varphi}-\frac{\cos(\theta)}{r^2 \sin^2(\theta)}\frac{\partial A_1}{\partial\varphi}\right\},
\end{eqnarray}
which is the induced static gravito-magnetic monopole charge density. For the spatial effective 4D currents of gravito-magnetic monopoles, we obtain
\begin{equation}
J_i=0, \qquad {i=2,3,4},
\end{equation}
which is consistent with the magnetic charge conservation in a scenario in which the monopoles are static.
Now we must study some important invariants of the problem. We define
\begin{eqnarray}
 \label{inv00}^{(m)}\underrightarrow{\underrightarrow{J}}\left(\,^{(e)}\overrightarrow{J},\overrightarrow{U}\right)&=&\,^{(e)}J^{A}\,^{(m)}J_{AB}\,U^{B}=\,^{(m)}J_{A} \,^{(e)}J^{A},\\
 \label{inv01}^{(m)}J&:=&\,^{(m)}\overrightarrow{J}\left(\overrightarrow{U}\right)=\,^{(m)}J_{A}\,U^{A},\\
 \label{inv02}^{(e)}J&:=&\,^{(e)}\overrightarrow{J}\left(\overrightarrow{U}\right)=\,^{(e)}J_{A}\,U^{A},\\
 \label{inv03} ^{(gem)}J^2&:=&\left(^{ (e)}\underrightarrow{J} \, \wedge
\,^{(m)}\underrightarrow{J}\right)\left(^{(e)}\overrightarrow{J}\,\wedge\,^{(m)}\overrightarrow{J}\right)=\,^{(gem)}J_{AB}\,^{(gem)}J^{AB},
\end{eqnarray}
These expressions must be constructed as a function of the LC or
Weitzenb\"{o}ck currents. The $U^{B}$ are the components of the
penta-velocities of the observers, expressed in the RN
non-coordinate basis. Hence, all the invariants (\ref{inv00}),
(\ref{inv01}) (\ref{inv02}) and (\ref{inv03}), defined in 5D, are
constructed taking into account these penta-velocities and the
gravito-electric and gravito-magnetic currents. From the analysis of the
invariant (\ref{inv00}) we see that the current
characterized by expression $^{(m)}J_A:=\,^{(m)}J_{AB}\,U^{B}$ comes from the contraction between
the ($2$-form) magnetic current
$^{(m)}\underrightarrow{\underrightarrow{J}}$ and the ($1$-vector)
penta-velocity $\overrightarrow{U}$. Hence, the complete invariant (\ref{inv00}) must be interpreted as
an inner product
$^{(m)}\overrightarrow{J}\cdot\,^{(e)}\overrightarrow{J}=\,^{(m)}J_A\,^{(e)}J^A$,
which means that both involved currents must be orthogonal when (\ref{inv00}) is zero. On the other hand, the
invariant (\ref{inv03}) is a new
$2$-form, which contains information as regards gravity, electricity, and magnetism. It is due to the wedge product of the
($1$-form) magnetic current $^{(m)}\underrightarrow{J}$ and the
($1$-form) electric current $^{(e)}\underrightarrow{J}$. We call such a new $2$-form as
$^{(gem)}\underrightarrow{\underrightarrow{J}}$, in which
supra-index $^{(gem)}$ refers to the gravito-electro-magnetic
nature of the current. In general
$^{(gem)}J^{AB}=\,^{(e)}\overrightarrow{J}\,\wedge\,^{(m)}\overrightarrow{J}$
must be not null. The expression (\ref{inv03}) is equivalent to the
inner product of two $2$-forms, which are expressed by
$\left[^{(e)}\underrightarrow{J}\wedge
^{(m)}\underrightarrow{J}\right]_{AB}\left[^{(e)}\overrightarrow{J}\wedge
^{(m)}\overrightarrow{J}\right]^{AB}=\,^{(gem)}J^2$.

The
invariants $^{(e)}J$ and $^{(m)}J$ have direct counterparts in the
4D space-time, defined by
\begin{eqnarray}\label{inv014}
\left.^{(m)}J\right|_{4D}&=&\left.^{(m)}J_{A}\,U^{A}\right|_{fol},\\ \label{inv024}
\left.^{(e)}J\right|_{4D}&=&\left.^{(e)}J_{A}\,U^{A}\right|_{fol}.
\end{eqnarray}
On the other hand, the 4D counterpart of $^{(m)}J_{A} \,^{(e)}J^{A}$ is
\begin{eqnarray}\label{inv004}
\left.^{(m)}J_{A} \,^{(e)}J^{A} \right|_{4D}= \left.\,^{(m)}J_{A}\,^{(e)}J^{A}\right|_{fol}.
\end{eqnarray}
The invariant $^{(m)}J_{A}\, ^{(e)}J^{A}=0$ only makes sense because we are supposing that the 5-vectors $^{(m)}J$ and $^{(e)}J$ are orthogonal on the 5D Weitzenb\"ock vacuum.
All the previously defined invariants have a clear meaning, which can easily be explored in a static or stationary case, as the case of a RN BH, which we shall study.
Then we must show that expressions (\ref{inv00}-\ref{inv004}) reduce to \begin{footnote}{In a static 5D space-time in which the observer is moving with arbitrary penta-velocity $\overrightarrow{U}=U^A\overrightarrow{E}_A$, their components $U^A$ can be $U^1,U^i,U^5\neq 0$. Hence, one obtains
\begin{eqnarray}\nonumber
[^{(m)}J_{AB}]=\left(
                                                \begin{array}{ccccc}
                                                  0 & 0 & 0 & 0 & -\rho_m \\
                                                  0 & 0 & 0 & 0 & -\,^{(m)}J_2 \\
                                                  0 & 0 & 0 & 0 & -\,^{(m)}J_3 \\
                                                  0 & 0 & 0 & 0 & -\,^{(m)}J_4 \\
                                                  \rho_m & \,^{(m)}J_2 & \,^{(m)}J_3 & \,^{(m)}J_4 & 0 \\
                                                \end{array}
                                              \right) \longrightarrow \left(
                                                \begin{array}{ccccc}
                                                  0 & 0 & 0 & 0 & -\rho_m \\
                                                  0 & 0 & 0 & 0 & 0 \\
                                                  0 & 0 & 0 & 0 & 0 \\
                                                  0 & 0 & 0 & 0 & 0 \\
                                                  \rho_m & 0 & 0 & 0 & 0 \\
                                                \end{array}
                                              \right),
\end{eqnarray}
in which we have made use of the fact that $^{(m)}J_i=0$, so that
\begin{eqnarray}\nonumber
[^{(m)}J_{A}]=(-\rho_m \, U^5,0,0,0,\rho_m \, U^1).
\end{eqnarray}
In a non-static scenario results that $^{(m)}J_5=0$. On the other hand, using the fact that $^{(e)}J_i=0$, we obtain
\begin{eqnarray}\nonumber
[^{(e)}J_{A}]=(\rho_e,0,0,0,\,^{(e)}J_5),
\end{eqnarray}
where in general $^{(e)}J_5\neq 0$. All makes sense when we apply our extended 5D Maxwell equations for $^{(e)}J_5$, so that ${F_{5A}}^{;\,A}=\,^{(e)} J_5$, which in the
extended version is
\begin{eqnarray}\nonumber
^{(e)}J_5=-\Box \Psi + {\left(A_{A}^{;\,A}\right)}_{;\,5}.
\end{eqnarray}
If we apply the 5D Lorenz gauge ($A_A^{;\,A}=0$), the last expression evaluated on the foliation reduces to
\begin{eqnarray}\label{j5}
^{(e)}J_5\mid_{fol}=-\Box \Psi \mid_{fol}.
\end{eqnarray}
The expression (\ref{j5}) must be interpreted as an effective equation of the form $\rho_M=-\Box \Psi$, in which $\Psi$ is the gravitational potential for the 4D BH.
In this case $^{(e)}J_5$ must be $\rho_M$; the effective density of the gravitational mass.
We must remark that in the present case the operator $\Box$ does not have a time related covariant derivative part, due to the static nature of the problem here considered (we maintain the complete notation for clarity).
We must set
\begin{eqnarray}\label{je5}
[^{(e)}J_{A}]=(\rho_e,0,0,0,\rho_M).
\end{eqnarray}
The Equation (\ref{je5}) justifies the notation used in Eqs. (\ref{1}) and (\ref{inv0245}).}\end{footnote}
\begin{eqnarray}\label{1}
\left.^{(m)}J_{A}\, ^{(e)}J^{A}\right|_{fol}&=&\left.\rho_m\rho_M U^1+\rho_e\rho_m U^5\right|_{fol}=0,\\\label{2}
\left.^{(m)}J_{A}  U^A\right|_{fol}&=&\left.\rho_m\,\left(U^1 U^5 - U^5 U^1\right)\right|_{fol}=0,\\
\label{inv0245}
\left.^{(e)}J_{A}  U^A\right|_{fol}&=&\left.\rho_e\,U^1+\rho_{M}U^5\right|_{fol}=0. \label{inv0246}
\end{eqnarray}
We shall study the case with $U^i=0$ and $U^5 \neq 0$.
For such scenario we assume that the observer is moving with a penta-velocity $\overrightarrow{U}=U^A\overrightarrow{E}_A$, with $U^A\neq 0$. Then we must set $U^5$ to an arbitrary non-zero value. The penta-velocity is defined on a 5D TM and is a penta-vector\footnote{The notation used is: \begin{itemize}\item The 5D M is the 5D manifold.\item The 5D TM is tangent space associated to 5D M.\item 4D M is the 4D manifold induced by foliation $\psi=\psi_0$ and, of course, a sub-manifold of the 5D M.\item The 4D TM is the tangent space associated to 4D M and, of course, a sub-space of the 5D TM.\end{itemize}}.

When we apply a constant foliation $\psi=\psi_0$ on the extra
coordinate, the penta-velocity turns in a four-vector by
projecting it onto the 4D TM. Then
$U^1\overrightarrow{E}_1+U^i\overrightarrow{E}_i+U^5\overrightarrow{E}_5\rightarrow
u^1\overrightarrow{e}_1+u^i\overrightarrow{e}_i\mid_{\psi=\psi_0}$,
so that
$\{\overrightarrow{E}_1,\overrightarrow{E}_i,\overrightarrow{E}_5\}$
is a basis of the 5D TM, and
$\{\overrightarrow{e}_1,\overrightarrow{e}_i\}$ is a basis of the
induced 4D TM. Due to the static nature of the studied scenario
the invariants (\ref{1}-\ref{inv0246}) are not dependent of $u^i$
and we must choose the value of $u^i$ preserving the validity of
the geodesic equations: $\frac{du^{\alpha}}{dS} +
^{(W)}\Gamma^{\alpha}_{\beta\gamma} u^{\beta} u^{\gamma}=0$, with
the normalization condition
$g_{\alpha\beta}u^{\alpha}u^{\beta}=1$. Here, $g_{\alpha\beta}$ is
the effective induced metric with the foliation $\psi=\psi_0$ and
we must take the RN, or the GP ordinary 4D form, depending on the
employed coordinates.

The use of WIMT warrants that the 5D space-time is a vacuum in
the sense of the Weitzenb\"{o}ck geometry, even if
$\rho_m,\,\rho_e,\,\rho_M\neq 0$. At this point we choose to
establish some values for the 5D invariants, in order to obtain
a realistic physical behavior in the effective space-time. Hence,
we set $^{(m)}J_{A} \,^{(e)}J^{A}=0$ to find
\begin{eqnarray}\label{3}
\frac{\rho_M}{\rho_e}=-\frac{U^5}{U^1}.
\end{eqnarray}

On the another hand, from $^{(e)}J=0$, we obtain that
\begin{equation}\label{4}
\frac{\rho_e}{\rho_M} = -\frac{U^5}{U^1},
\end{equation}
so that, equating (\ref{3}) and (\ref{4}), we obtain that
\begin{equation}\label{5}
\frac{U^5}{U^1} =\frac{\rho_M}{\rho_e} = \pm 1,
\end{equation}
which provides us with the quantization law between mass density and the
electric density charge. In order to obtain the quantization of
charge we must use the Eqs. (\ref{inv03}) and (\ref{5}),
with $^{(gem)}J^2=-n^2$. The resulting expression, which provides
the quantization between the electric and magnetic density charges,
is
\begin{equation}\label{qu}
\rho_e\,\rho_m = \frac{n}{2}.
\end{equation}
Notice that all these expressions were obtained in the ONB representation. A very important fact is that the choices $^{(m)}J_{A}\, ^{(e)}J^{A}=0$ and
$^{(gem)}J^2=-n^2$ imply that the duality of charges $\rho_e$ and $\rho_m$ are mutually exclusive, with the constriction (\ref{qu}). This means that one only can "see" the electric $\rho_e$ or the magnetic charge
$\rho_m = {n\over 2 \rho_e}$. This is because Eq. (\ref{2}) indicates that $^{(m)} \vec{J}$ and $^{(e)} \vec{J}$ are orthogonal. Therefore, our model with these choices is incompatible with a dyonic Reissner-Nordstr\"{o}m BH, except for $n=0$. In order to
make our theory compatible with a dyonic Reissner-Nordstr\"{o}m BH one should consider that $^{(m)}J_{A}\, ^{(e)}J^{A}\neq 0$ and $^{(gem)}J^2=-n^2$, but this issue will be studied in a future work.

We shall return to the expression (\ref{current0}), in order to see the consistency of such an equation for the 4D Wu-Yang potentials, which describe a localized magnetic monopole with charge $q_m$ \cite{wu}
\begin{eqnarray}\label{pot1}
\overrightarrow{A}^{(N)}&=&q_m \frac{(1-\cos(\theta))}{r \sin(\theta)}\overrightarrow{e}_\varphi=\frac{q_m}{r}A^{(N)}_3\overrightarrow{e}_{\varphi},\\ \label{pot2}
\overrightarrow{A}^{(S)}&=& -q_m \frac{\left(1+\cos(\theta)\right)}{r \sin(\theta)}\overrightarrow{e}_\varphi=\frac{q_m}{r}A^{(S)}_3\overrightarrow{e}_{\varphi}.
\end{eqnarray}
Here, labels $(N)$ and $(S)$ indicate the North or South hemisphere, on which are valid the potentials. The magnetic monopole density in the
RN metric is
\begin{eqnarray}\label{ro-}
\rho^{(*)}_m=-\frac{K}{r}\frac{\partial A^{(*)}_3}{\partial\theta},
\end{eqnarray}
in which the asterisk $(*)$ stands for the labels $(N)$ or $(S)$, depending on what hemisphere we are dealing with. For simplicity we choose to integrate the last expression in the
orthonormalized metric, in which the associated volume element is $d\,V=\frac{1}{\sqrt{f(r)}}\,r^2 \,\sin(\theta) \,d\theta\,d\varphi\,dr$ due to $\mid g\mid=1$ \begin{footnote}{In the orthonormalized basis $g_{\alpha\beta}=\eta_{\alpha\beta}=\pm 1$. Then $\mid g\mid =1$ and the basis of the cotangent space is $\underrightarrow{E}_1=\sqrt{f(r)}\underrightarrow{dt},\,\underrightarrow{E}_2=\frac{1}{\sqrt{f(r)}}\underrightarrow{dr},\,\underrightarrow{E}_3=r \underrightarrow{d\theta},\,\underrightarrow{E}_4=r\sin(\theta)\underrightarrow{d\varphi}$}. Therefore
\begin{eqnarray}\nonumber d\,V=\, \frac{1}{\sqrt{f(r)}}\,r^2\, \sin(\theta)\,d\theta\,d\varphi\,dr.
\end{eqnarray}
On the other hand, the current is
\begin{eqnarray}\nonumber
^{(m)}\overrightarrow{J}&=&\underbrace{^{(m)}J(ON)_0}\frac{1}{\sqrt{f(r)}}\overrightarrow{\frac{\partial}{\partial\,t}}\Rightarrow
\\&\,&\nonumber orthonormalized\,basis\\\nonumber
\underbrace{^{(m)}J(RN)_0}&=&\frac{1}{\sqrt{f(r)}}\,^{(m)}J(ON)_0.\\\nonumber
RN basis,
\end{eqnarray}
Therefore, by integrating on the North and South hemispheres we find that the magnetic charge in the BH is given by
\begin{eqnarray}\nonumber
q^{(*)}_m=-K\int\int\int_{(*)} \frac{\partial\,A^{(*)}_3}{\partial\,\theta}\sin(\theta)\,d\theta\,d\varphi\,dr,
\end{eqnarray}
in which $q^{(*)}_m$ is the total magnetic charge on the corresponding hemisphere. Of course, the total BH magnetic charge will be $q_m=q^{(N)}_m+q^{(S)}_m$.\end{footnote},
and the integral on the volume for the total magnetic charge of the BH takes the form
\begin{eqnarray}\label{qm} &\,&\\ \nonumber
q_m &=&\int\int\int^{(N)}\rho^{(N)}_m \,d\,V+\int\int\int^{(S)}\rho^{(S)}_m \,d\,V=\\\nonumber&=&-K\left(\int\int\int^{(N)}\frac{\partial A^{(N)}_3}{\partial\theta}\sin(\theta)\, d\theta\,d\varphi\,dr+\int\int\int^{(S)}\frac{\partial A^{(S)}_3}{\partial\theta}\sin(\theta)\, d\theta\,d\varphi\,dr\right),
\end{eqnarray}
which leads to
\begin{eqnarray}\label{carga}
\int\int\int_{BH}\rho_m \,d\,V=-2\pi\,K\,r_{int}q_m,
\end{eqnarray}
$r_{int}=M-\sqrt{M^2-Q^2}$ being the interior radius of the RN BH.
We assume that $Q<M$, so that for consistence we shall require that $K=-\frac{1}{2\pi \,r_{int}}$.
Then to the effective RN BH, in which our WIMT induces a gravito-magnetic current, belongs a  density of monopole charge $\rho_m$, which is not uniform ($\rho_m\neq\frac{q_m}{\frac{4}{3}\pi{r_{int}}^3}$), but compatible with a total magnetic charge $q_m$, over the BH, which is the origin of the Wu-Yang potentials.

\section{Final remarks}

We have employed WIMT over the 5D GP and RN metrics in order to obtain the Weitzenb\"{o}ck torsion in the ONB representation, which must be
transformed with the vielbein (\ref{egp01}) and (\ref{egp02}) to the GP representation. The matrix which represents the vielbein has an antisymmetric part,
with which we obtained an effective non-zero gravito-magnetic current originating in the Weitzenb\"{o}ck torsion. The effective current in the RN representation gives rise to a non-zero magnetic monopole density without any time dependence. We have confirmed that it is compatible with the Wu-Yang potentials for a localized magnetic charge of magnitude $q_m$. In our case the magnetic density $\rho_m$ must be seen as a magnetic charge distribution over the BH, where the $\theta$-dependence (which is a manifestation of an anisotropic distribution of charge), represents the total magnetic charge $q_m$, of the BH. We analyze various invariants which make it possible to interpret the physical magnitudes of our problem $\left(\rho_m, \rho_M, \rho_e\right)$ as elements coming from such invariants; this highlights the correspondence between such magnitudes and the geometry of the space-time. In particular, the quantization condition arises as an effective manifestation of the current expression between the components of the penta-velocity of the observer. Finally, we have obtained an induced gravito-magnetic monopole in the framework of WIMT, which is compatible with the Wu-Yang fields, with the same magnitude of charge and quantization of charge included. This is a very important result that manifests the power of our formalism.

\appendix

\section{Conecting ONB, GP and RN.}

Here we try to order the different representations of space-time and the relationship between them:
\begin{itemize}
\item {\bf ONB representation}: in 5D the orthonormal basis of 5D TM is $\{\overrightarrow{E}_A\}$, which is non-holonomic with structure coefficients given in (\ref{egp04}) and the subsequent expressions. In this case the matrix of the metric tensor is $\eta_{AB}$ with $\overrightarrow{E}_A\cdot\overrightarrow{E}_B=\eta_{AB}=\pm 1$. With a constant foliation we must obtain $\eta_{\alpha\beta}=\pm 1$, for the basis $\{\overrightarrow{E}_\alpha\}$. In this context the Weitzenb\"{o}ck torsion is specially easy to obtain according to (\ref{egp05}), and we can transform it to other representations using the corresponding vielbein.
\item {\bf GP representation}: The 5D GP cartesian metric (\ref{egp03}) is linked to the coordinate basis $\overrightarrow{e}_a=\overrightarrow{\frac{\partial\,\,}{\partial \,x^a}}$ associated to $x^a$, with $x^1=t_{GP}$, $x^2=x$, $x^3=y$, $x^4=z$, $x^5=l$, in which $t_{GP}=t+h(r)$, where $h(r)$ is an arbitrary ($C^{\infty}$) function. Using a constant foliation we must obtain the familiar 4D GP cartesian metric \cite{gpCASA}.
\item {\bf RN representation}: The usual 5D RN metric associated to an effective 4D charged black hole which in 5D is characterized by a length element $dS^2=f(r)dt^2-{f(r)}^{-1}dr^2-d\Omega^2-dl^2$, with $f(r)=1-\frac{2M}{r}+\frac{Q^2}{r^2}$ expressed for the coordinate basis $\{\overrightarrow{e}'_{a'}=\overrightarrow{\frac{\partial\,\,}{\partial\,x'^{a'}}}\}$, associated to $x'^{a'}$, where $x'^1=t_{RN}$, $x'^2=r$, $x'^3=\theta$,
    $x'^4=\varphi$, $x'^5=l$. The 4D RN metric must be obtained by a constant foliation.
\item {\bf Tensorial objects}: in the ONB representation must be transformed to the GP representation using the vielbein (\ref{egp01}) and (\ref{egp02}), and vice versa.

To obtain the vielbein connecting GP and RN, we must use the fact that both bases are coordinate (holonomic). Then $\overrightarrow{e}'_{a'}=e_{a'}^a\overrightarrow{e}_a$ with $e_{a'}^a=\frac{\partial x^a}{\partial x'^{a'}}$ and vice versa. The non-zero vielbeins relating the ONB with RN, according to $\overrightarrow{E}_A=e^{a'}_A\overrightarrow{e}_{a'}$, are $e^{a'=1}_{A=1}=\frac{1}{\sqrt{f(r)}}$, $e^{a'=2}_{A=2}=\sqrt{f(r)}$, e$^{a'=3}_{A=3}=\frac{1}{r}$, $e^{a'=4}_{A=4}=\frac{1}{r\sin(\theta)}$, $e^{a'=5}_{A=5}=1$, and therefore they comply the expressions $e^{a'}_A e^A_{b'}=\delta^{a'}_{b'}$ and $e^{a'}_Ae^B_{a'}=\delta^{B}_{A}$.
   \item {\bf Constant foliation}: must be viewed as the application of a vielbein $e^a_\alpha=\delta^a_\alpha$, $e^5_\alpha=0$ to the tensor objects plus a specialization on certain value for $\psi=\psi_0$. Then, for example, $F^a_b\rightarrow F^{\alpha}_{\beta}=e^{\alpha}_a e_{\beta}^b F^{a}_b\mid_{\psi=\psi_0}$. This idea must be extended to another more general kind of foliation (dynamic foliation) varying the vielbein as to be required.
\end{itemize}
Using the previous concepts and formulas we must relate ONB, GP, and RN representations and the effective 4D space-time, tensor objects, and invariants.
\section*{Acknowledgements}

\noindent J. M. Romero and M. Bellini acknowledge CONICET
(Argentina) and UNMdP for financial support.

\end{document}